\documentclass[italian,english]{article}
\usepackage[T1]{fontenc}
\usepackage[latin1]{inputenc}
\usepackage{color}
\usepackage{amssymb}

\makeatletter

 \newcommand{\lyxaddress}[1]{
   \par {\raggedright #1 
   \vspace{1.4em}
   \noindent\par}
 }

\usepackage{babel}
\makeatother
\begin{document}

\title{\textbf{An oscillating Universe from the linearized $R^{2}$ theory
of gravity }}

\author{\textbf{Christian Corda}}

\maketitle

\lyxaddress{\begin{center}Associazione Galileo Galilei, Via Pier Cironi 16 -
59100 PRATO, Italy; Dipartimento di Fisica Università di Pisa, Largo
Pontecorvo 3 - 56127 PISA, Italy; \end{center}}

\lyxaddress{\begin{center}\textit{E-mail address:} \textcolor{blue}{christian.corda@ego-gw.it}\end{center}}

\begin{abstract}
An oscillating Universe which arises from the linearized \textbf{$R^{2}$}
theory of gravity is discussed, showing that some observative evidences
like the cosmological redshift and the Hubble law are in agreement
with the model. In this context Dark Energy is seen like a pure curvature
effect arising by the Ricci scalar.
\end{abstract}

\lyxaddress{PACS numbers: 04.80.Nn, 04.30.Nk, 04.50.+h}

\section{Introduction}

The accelerated expansion of the Universe, which is today observed,
shows that cosmological dynamic is dominated by the so called Dark
Energy which gives a large negative pressure. This is the standard
picture, in which such new ingredient is considered as a source of
the \textit{rhs} of the field equations. It should be some form of
un-clustered non-zero vacuum energy which, together with the clustered
Dark Matter, drives the global dynamics. This is the so called {}``concordance
model'' (ACDM) which gives, in agreement with the CMBR, LSS and SNeIa
data, a good trapestry of the today observed Universe, but presents
several shortcomings as the well known {}``coincidence'' and {}``cosmological
constant'' problems \cite{key-1}. 

An alternative approach is changing the \textit{lhs} of the field
equations, seeing if observed cosmic dynamics can be achieved extending
general relativity \cite{key-2,key-3,key-4}. In this different context,
it is not required to find out candidates for Dark Energy and Dark
Matter, that, till now, have not been found, but only the {}``observed''
ingredients, which are curvature and baryonic matter, have to be taken
into account. Considering this point of view, one can think that gravity
is not scale-invariant \cite{key-5} and a room for alternative theories
is present \cite{key-6,key-7,key-8}. In principle, the most popular
Dark Energy and Dark Matter models can be achieved considering $f(R)$
theories of gravity \cite{key-5,key-9}, where $R$ is the Ricci curvature
scalar. In this picture even the sensitive detectors for gravitational
waves, like bars and interferometers (i.e. those which are currently
in operation and the ones which are in a phase of planning and proposal
stages) \cite{key-10,key-11}, could, in principle, be important to
confirm or ruling out the physical consistency of general relativity
or of any other theory of gravitation. This is because, in the context
of Extended Theories of Gravity, some differences between General
Relativity and the others theories can be pointed out starting by
the linearized theory of gravity \cite{key-12,key-13,key-14,key-15}. 

In this paper an oscillating Universe which arises by the linearized
\textbf{$R^{2}$} theory of gravity, proposed the first time by Starobinsky
in \cite{key-2}, which is the simplest among the $f(R)$ theories
of gravity \cite{key-3,key-4}, is discussed, showing that some observative
evidences, like the cosmological redshift and the Hubble law, are
in agreement with the model. In this context Dark Energy is seen like
a pure curvature effect arising by the Ricci scalar. The paper is
organized in this way: in the second section it is shown that a third
mode of gravitational radiation arises from the high order action
of the Starobinsky's theory of gravity\begin{equation}
S=\int d^{4}x\sqrt{-g}(R+\alpha R^{2})+\mathcal{L}_{m}.\label{eq: high order 1}\end{equation}

Equation (\ref{eq: high order 1}) is a particular choice with respect
the well known canonical one of general relativity (the Einstein -
Hilbert action \cite{key-16,key-17}) which is 

\begin{equation}
S=\int d^{4}x\sqrt{-g}R+\mathcal{L}_{m}.\label{eq: EH}\end{equation}

In the third section, with the assumption that this third mode becomes
dominant at cosmological scales, an oscillating model of Universe
will be shown.

Finally, in the fourth section, the tuning with observations will
be analysed.

\section{The linearized $R^{2}$ theory of gravity}

If the gravitational Lagrangian is nonlinear in the curvature invariants,
the Einstein field equations has an order higher than second \cite{key-6,key-12,key-13}.
For this reason such theories are often called higher-order gravitational
theories. This is exactly the case of the action (\ref{eq: high order 1}).

Note that in this paper we work with $8\pi G=1$, $c=1$ and $\hbar=1,$
while the sign conventions for the line element, which generate the
sign conventions for the Riemann/Ricci tensors, are (+,-,-,- ). 

By varying the action (\ref{eq: high order 1}) with respect to $g_{\mu\nu}$
(see refs. \cite{key-6,key-12,key-13} for a parallel computation)
the field equations are obtained:

\begin{equation}
\begin{array}{c}
G_{\mu\nu}+\alpha\{2R[R_{\mu\nu}-\frac{1}{4}g_{\mu\nu}R]+\\
\\-2R_{;\mu;\nu}+2g_{\mu\nu}\square R\}=T_{\mu\nu}^{(m)}\end{array}\label{eq: einstein-general}\end{equation}

with associed a Klein - Gordon equation for the Ricci curvature scalar 

\begin{equation}
\square R=E^{2}(R+T),\label{eq: KG}\end{equation}

which is obtained by taking the trace of equation (\ref{eq: einstein-general}),
where $\square$ is the d' Alembertian operator and the energy $E$
has been introduced for dimensional motivations: $E^{2}\equiv\frac{1}{6\alpha}$,
thus $\alpha$ has to be positive \cite{key-2}.

In the above equations $T_{\mu\nu}^{(m)}$ is the ordinary stress-energy
tensor of the matter. Note that General Relativity is obtained for
$\alpha=0.$

Because we want to study interactions at cosmological scales, the
linearized theory in vacuum ($T_{\mu\nu}^{(m)}=0$), which gives a
better approximation than Newtonian theory, can be analyzed, with
a little perturbation of the background, which is assumed given by
a Minkowskian background. 

Putting

\begin{equation}
g_{\mu\nu}=\eta_{\mu\nu}+h_{\mu\nu}\label{eq: linearizza}\end{equation}

\begin{equation}
\frac{R}{E}\simeq0+\frac{\delta R}{E}\equiv h_{R},\label{eq: linearicci}\end{equation}

to first order in $h_{\mu\nu}$ , calling $\widetilde{R}_{\mu\nu\rho\sigma}$
, $\widetilde{R}_{\mu\nu}$ and $\widetilde{R}$ the linearized quantity
which correspond to $R_{\mu\nu\rho\sigma}$ , $R_{\mu\nu}$ and $R$,
the linearized field equations are obtained \cite{key-12,key-13,key-16,key-17}:

\begin{equation}
\begin{array}{c}
\widetilde{R}_{\mu\nu}-\frac{\widetilde{R}}{2}\eta_{\mu\nu}=-\partial_{\mu}\partial_{\nu}h_{R}+\eta_{\mu\nu}\square h_{R}\\
\\{}\square h_{R}=E^{2}h_{R}.\end{array}\label{eq: linearizzate1}\end{equation}

$\widetilde{R}_{\mu\nu\rho\sigma}$ and eqs. (\ref{eq: linearizzate1})
are invariants for gauge transformations \cite{key-6,key-8}

\begin{equation}
\begin{array}{c}
h_{\mu\nu}\rightarrow h'_{\mu\nu}=h_{\mu\nu}-\partial_{(\mu}\epsilon_{\nu)}\\
\\h_{R}\rightarrow h_{R}'=h_{R};\end{array}\label{eq: gauge}\end{equation}

then 

\begin{equation}
\bar{h}_{\mu\nu}\equiv h_{\mu\nu}-\frac{h}{2}\eta_{\mu\nu}+\eta_{\mu\nu}h_{R}\label{eq: ridefiniz}\end{equation}

can be defined, and, considering the transform for the parameter $\epsilon^{\mu}$

\begin{equation}
\square\epsilon_{\nu}=\partial^{\mu}\bar{h}_{\mu\nu},\label{eq:lorentziana}\end{equation}
 a gauge parallel to the Lorenz one of electromagnetic waves can be
choosen:

\begin{equation}
\partial^{\mu}\bar{h}_{\mu\nu}=0.\label{eq: cond lorentz}\end{equation}

In this way field equations read like

\begin{equation}
\square\bar{h}_{\mu\nu}=0\label{eq: onda T}\end{equation}

\begin{equation}
\square h_{R}=E^{2}h_{R}\label{eq: onda S}\end{equation}

Solutions of eqs. (\ref{eq: onda T}) and (\ref{eq: onda S}) are
plan waves:

\begin{equation}
\bar{h}_{\mu\nu}=A_{\mu\nu}(\overrightarrow{p})\exp(ip^{\alpha}x_{\alpha})+c.c.\label{eq: sol T}\end{equation}

\begin{equation}
h_{R}=a(\overrightarrow{p})\exp(iq^{\alpha}x_{\alpha})+c.c.\label{eq: sol S}\end{equation}

where

\begin{equation}
\begin{array}{ccc}
k^{\alpha}\equiv(\omega,\overrightarrow{p}) &  & \omega=p\equiv|\overrightarrow{p}|\\
\\q^{\alpha}\equiv(\omega_{E},\overrightarrow{p}) &  & \omega_{E}=\sqrt{E^{2}+p^{2}}.\end{array}\label{eq: k e q}\end{equation}

In eqs. (\ref{eq: onda T}) and (\ref{eq: sol T}) the equation and
the solution for the tensorial waves exactly like in general relativity
have been obtained \cite{key-16,key-17}, while eqs. (\ref{eq: onda S})
and (\ref{eq: sol S}) are respectively the equation and the solution
for the third mode arising by curvature (see also \cite{key-12,key-13}).

The fact that the dispersion law for the modes of the field $h_{R}$
is not linear has to be emphatized. The velocity of every {}``ordinary''
(i.e. which arises from General Relativity) mode $\bar{h}_{\mu\nu}$
is the light speed $c$, but the dispersion law (the second of eq.
(\ref{eq: k e q})) for the modes of $h_{R}$ is that of a wave-packet
\cite{key-12,key-13}. Also, the group-velocity of a wave-packet of
$h_{R}$ centered in $\overrightarrow{p}$ is 

\begin{equation}
\overrightarrow{v_{G}}=\frac{\overrightarrow{p}}{\omega}.\label{eq: velocita' di gruppo}\end{equation}

From the second of eqs. (\ref{eq: k e q}) and eq. (\ref{eq: velocita' di gruppo})
it is simple to obtain:

\begin{equation}
v_{G}=\frac{\sqrt{\omega^{2}-E^{2}}}{\omega}.\label{eq: velocita' di gruppo 2}\end{equation}

Then, it is also \cite{key-12,key-13}

\begin{equation}
E=\sqrt{(1-v_{G}^{2})}\omega.\label{eq: relazione massa-frequenza}\end{equation}

Now the analisys can remain in the Lorenz gauge with trasformations
of the type $\square\epsilon_{\nu}=0$; this gauge gives a condition
of transversality for the ordinary part of the field: $k^{\mu}A_{\mu\nu}=0$,
but does not give the transversality for the total field $h_{\mu\nu}$.
From eq. (\ref{eq: ridefiniz}) it is

\begin{equation}
h_{\mu\nu}=\bar{h}_{\mu\nu}-\frac{\bar{h}}{2}\eta_{\mu\nu}+\eta_{\mu\nu}h_{m}.\label{eq: ridefiniz 2}\end{equation}

At this point, if beeing in general relativity, one puts \cite{key-12,key-17},

\begin{equation}
\begin{array}{c}
\square\epsilon^{\mu}=0\\
\\\partial_{\mu}\epsilon^{\mu}=-\frac{\bar{h}}{2}+h_{R},\end{array}\label{eq: gauge2}\end{equation}

which gives the total transversality of the field. But, in the present
case, this is impossible. In fact, applying the d'Alembertian operator
to the second of eqs. (\ref{eq: gauge2}) and using the field equations
(\ref{eq: onda T}) and (\ref{eq: onda S}), it results

\begin{equation}
\square\epsilon^{\mu}=E^{2}h_{R},\label{eq: contrasto}\end{equation}

which is in contrast with the first of eqs. (\ref{eq: gauge2}). In
the same way it is possible to show that it does not exist any linear
relation between the tensorial field $\bar{h}_{\mu\nu}$ and the {}``curvature''
field $h_{R}$. Thus a gauge in wich $h_{\mu\nu}$ is purely spatial
cannot be chosen (i.e. it cannot be put $h_{\mu0}=0,$ see eq. (\ref{eq: ridefiniz 2}))
. But the traceless condition to the field $\bar{h}_{\mu\nu}$ can
be put :

\begin{equation}
\begin{array}{c}
\square\epsilon^{\mu}=0\\
\\\partial_{\mu}\epsilon^{\mu}=-\frac{\bar{h}}{2}.\end{array}\label{eq: gauge traceless}\end{equation}

These equations imply

\begin{equation}
\partial^{\mu}\bar{h}_{\mu\nu}=0.\label{eq: vincolo}\end{equation}

To save the conditions $\partial_{\mu}\bar{h}^{\mu\nu}$ and $\bar{h}=0$
transformations like

\begin{equation}
\begin{array}{c}
\partial_{\mu}\square\epsilon^{\mu}=0\\
\\\partial_{\mu}\epsilon^{\mu}=0\end{array}\label{eq: gauge 3}\end{equation}

can be used and, taking $\overrightarrow{p}$ in the $z$ direction,
a gauge in which only $A_{11}$, $A_{22}$, and $A_{12}=A_{21}$ are
different to zero can be chosen. The condition $\bar{h}=0$ gives
$A_{11}=-A_{22}$. Now, putting these equations in eq. (\ref{eq: ridefiniz 2}),
it results

\begin{equation}
h_{\mu\nu}(t,z)=A^{+}(t-z)e_{\mu\nu}^{(+)}+A^{\times}(t-z)e_{\mu\nu}^{(\times)}+h_{R}(t,z)\eta_{\mu\nu}.\label{eq: perturbazione totale}\end{equation}

The term $A^{+}(t-z)e_{\mu\nu}^{(+)}+A^{\times}(t-z)e_{\mu\nu}^{(\times)}$
describes the two standard polarizations of gravitational waves which
arise from general relativity, while the term $h_{R}(t,z)\eta_{\mu\nu}$
is the {}``curvature'' field arising from the Starobinsky's high
order theory. In other words, in the \textbf{$R^{2}$} theory of gravity,
the Ricci scalar generates a third polarization for gravitational
waves which is not present in standard general relativity. This third
mode is associated to a {}``curvature'' energy $E$ (see equation
(\ref{eq: KG})).

\section{An oscillating Universe}

By assuming that, at cosmological scales, the third mode becomes dominant
(i.e. $A^{+},A^{-}\ll h_{R}$), as it appears from observations (i.e.
we are assuming that the {}``curvature energy'' is the Dark Energy
of the Universe $\simeq10^{-29}g/cm^{3}$ \cite{key-21,key-22}),
eq. (\ref{eq: perturbazione totale}) can be rewritten as

\begin{equation}
h_{\mu\nu}(t,z)=h_{R}(t,z)\eta_{\mu\nu}\label{eq: perturbazione scalare}\end{equation}
and the corrispondent line element is the conformally flat one

\begin{equation}
ds^{2}=[1+h_{R}(t,z)](dt^{2}-dz^{2}-dx^{2}-dy^{2}).\label{eq: metrica puramente scalare}\end{equation}

Defining 

\begin{equation}
a^{2}\equiv1+h_{R}(t,z),\label{eq: a quadro}\end{equation}

equation (\ref{eq: metrica puramente scalare}) becomes similar to
the well known cosmological Friedmann- Robertson Walker (FRW) line
element of the standard homogeneus and isotropic flat Universe which
is well known in the literature \cite{key-16,key-17,key-18,key-19,key-20}:

\begin{equation}
ds^{2}=[a^{2}(t,z)](+dt^{2}-dz^{2}-dx^{2}-dy^{2}).\label{eq: metrica FRW}\end{equation}
In our linearized approach it is also \begin{equation}
a\simeq1+\frac{1}{2}h_{R}(t,z),\label{eq: a}\end{equation}

which shows that in our model the scale factor of the Universe oscillates
near the (normalized) unity.

Below, it will be shown that our model realizes an oscillating homogeneus
and isotropic Universe, but, before starting with the analysis, we
have to recall that observations today agrees with homogeneity and
isotropy.

In Cosmology, the Universe is seen like a dynamic and thermodynamic
system in which test masses (i.e. the {}``particles'') are the galaxies
that are stellar systems with a number of the order of $10^{9}-10^{11}$
stars \cite{key-16,key-17,key-18,key-19,key-20}. Galaxies are located
in clusters and super clusters, and observations show that, on cosmological
scales, their distribution is uniform. This is also confirmed by the
WMAP data on the Cosmic Background Radiation \cite{key-21,key-22}.
These assumption can be summarized in the so called Cosmological Principle:
\textit{the Universe is homogeneous everyway and isotropic around
every point.} Cosmologic Principle semplifies the analysis of the
large scale structure, because it implies that the proper distances
between any two galaxies is given by an universal scale factor which
is the same for any couple of galaxies \cite{key-16,key-17,key-18,key-19,key-20}.

Because our observations are performed on Earth, the coordinate system
in which the space-time is locally flat has to be used and the distance
between any two points is given simply by the difference in their
coordinates in the sense of Newtonian physics \cite{key-12,key-13,key-14,key-15,key-16,key-17}.
This frame is the proper reference frame of a local observer, which
we assume to be located on Earth. In this frame gravitational signals
manifest themself by exerting tidal forces on the test masses, which
are the galaxies of the Universe. A detailed analysis of the frame
of the local observer is given in ref. \cite{key-17}, sect. 13.6.
Here only the more important features of this coordinate system are
recalled:

the time coordinate $x_{0}$ is the proper time of the observer O;

spatial axes are centered in O;

in the special case of zero acceleration and zero rotation the spatial
coordinates $x_{j}$ are the proper distances along the axes and the
frame of the local observer reduces to a local Lorentz frame: in this
case the line element reads \cite{key-17}

\begin{equation}
ds^{2}=+(dx^{0})^{2}-\delta dx^{i}dx^{j}-O(|x^{j}|^{2})dx^{\alpha}dx^{\beta}.\label{eq: metrica local lorentz}\end{equation}

The effect of the gravitational force on test masses is described
by the equation

\begin{equation}
\ddot{x^{i}}=-\widetilde{R}_{0k0}^{i}x^{k},\label{eq: deviazione geodetiche}\end{equation}
which is the equation for geodesic deviation in this frame.

Thus, to study the effect of the third mode of the linearized \textbf{$R^{2}$}
theory of gravity on the galaxies, $\widetilde{R}_{0k0}^{i}$ has
to be computed in the proper reference frame of the Earth. But, because
the linearized Riemann tensor $\widetilde{R}_{\mu\nu\rho\sigma}$
is invariant under gauge transformations \cite{key-12,key-13,key-17},
it can be directly computed from eq. (\ref{eq: perturbazione scalare}). 

From \cite{key-17} it is:

\begin{equation}
\widetilde{R}_{\mu\nu\rho\sigma}=\frac{1}{2}\{\partial_{\mu}\partial_{\beta}h_{\alpha\nu}+\partial_{\nu}\partial_{\alpha}h_{\mu\beta}-\partial_{\alpha}\partial_{\beta}h_{\mu\nu}-\partial_{\mu}\partial_{\nu}h_{\alpha\beta}\},\label{eq: riemann lineare}\end{equation}

that, in the case eq. (\ref{eq: perturbazione scalare}), begins

\begin{equation}
\widetilde{R}_{0\gamma0}^{\alpha}=\frac{1}{2}\{\partial^{\alpha}\partial_{0}h_{R}\eta_{0\gamma}+\partial_{0}\partial_{\gamma}h_{R}\delta_{0}^{\alpha}-\partial^{\alpha}\partial_{\gamma}h_{R}\eta_{00}-\partial_{0}\partial_{0}h_{R}\delta_{\gamma}^{\alpha}\};\label{eq: riemann lin scalare}\end{equation}

the different elements are (only the non zero ones will be written):

\begin{equation}
\partial^{\alpha}\partial_{0}h_{R}\eta_{0\gamma}=\left\{ \begin{array}{ccc}
\partial_{t}^{2}h_{R} & for & \alpha=\gamma=0\\
\\-\partial_{z}\partial_{t}h_{R} & for & \alpha=3;\gamma=0\end{array}\right\} \label{eq: calcoli}\end{equation}

\begin{equation}
\partial_{0}\partial_{\gamma}h_{R}\delta_{0}^{\alpha}=\left\{ \begin{array}{ccc}
\partial_{t}^{2}h_{R} & for & \alpha=\gamma=0\\
\\\partial_{t}\partial_{z}h_{R} & for & \alpha=0;\gamma=3\end{array}\right\} \label{eq: calcoli2}\end{equation}

\begin{equation}
-\partial^{\alpha}\partial_{\gamma}h_{R}\eta_{00}=\partial^{\alpha}\partial_{\gamma}h_{R}=\left\{ \begin{array}{ccc}
-\partial_{t}^{2}h_{R} & for & \alpha=\gamma=0\\
\\\partial_{z}^{2}h_{R} & for & \alpha=\gamma=3\\
\\-\partial_{t}\partial_{z}h_{R} & for & \alpha=0;\gamma=3\\
\\\partial_{z}\partial_{t}h_{R} & for & \alpha=3;\gamma=0\end{array}\right\} \label{eq: calcoli3}\end{equation}

\begin{equation}
-\partial_{0}\partial_{0}h_{R}\delta_{\gamma}^{\alpha}=\begin{array}{ccc}
-\partial_{t}^{2}h_{R} & for & \alpha=\gamma\end{array}.\label{eq: calcoli4}\end{equation}

Now, putting these results in eq. (\ref{eq: riemann lin scalare}),
it results:

\begin{equation}
\begin{array}{c}
\widetilde{R}_{010}^{1}=-\frac{1}{2}\ddot{h}_{R}\\
\\\widetilde{R}_{010}^{2}=-\frac{1}{2}\ddot{h}_{R}\\
\\\widetilde{R}_{030}^{3}=\frac{1}{2}(\partial_{z}^{2}h_{R}-\partial_{t}^{2}h_{R}).\end{array}\label{eq: componenti riemann}\end{equation}

But, the assumption of homogenity and isotropy implies $\partial_{z}h_{R}=0,$
which also implies

\begin{equation}
\begin{array}{c}
\widetilde{R}_{010}^{1}=-\frac{1}{2}\ddot{h}_{R}\\
\\\widetilde{R}_{010}^{2}=-\frac{1}{2}\ddot{h}_{R}\\
\\\widetilde{R}_{030}^{3}=-\frac{1}{2}\ddot{h}_{R}\end{array}\label{eq: componenti riemann 2}\end{equation}

which show that the oscillations of the Universe are the same in any
direction. 

Infact, using eq. (\ref{eq: deviazione geodetiche}), it results

\begin{equation}
\ddot{x}=\frac{1}{2}\ddot{h}_{R}x,\label{eq: accelerazione mareale lungo x}\end{equation}

\begin{equation}
\ddot{y}=\frac{1}{2}\ddot{h}_{R}y\label{eq: accelerazione mareale lungo y}\end{equation}

and 

\begin{equation}
\ddot{z}=\frac{1}{2}\ddot{h}_{R}z,\label{eq: accelerazione mareale lungo z}\end{equation}

which are three perfectly symmetric oscillations.

\section{Tuning with observations}

\subsection{The Hubble law}

The expansion of the Universe arises from the observations of E Hubble
in 1929 \cite{key-16,key-17,key-18,key-19,key-20}. The Hubble law
states that, galaxies which are at a distance $D,$ drift away from
Earth with a velocity 

\begin{equation}
v=H_{0}D.\label{eq: legge Hubble}\end{equation}

The today's Hubble expansion rate is 

\begin{equation}
H_{0}=h_{100}\frac{100Km}{sec\times Mpc}=3.2\times10^{-18}\frac{h_{100}}{sec}.\label{eq: cost. Hubble}\end{equation}

A dimensionless factor $h_{100}$ is included, which now is just a
convenience (in the past it came from an uncertainty in the value
of $H_{0}$). From the WMAP data it is $h_{100}=0.72\pm0.05$ \cite{key-21,key-22}.

Calling $f$ the frequency of the {}``cosmologic'' gravitational
wave and assuming that $f\ll H_{0}$ (i.e. the gravitational wave
is {}``frozen'' with respect the cosmologic observations), the observations
of Hubble and the more recent ones imply that our model of oscillating
Universe has to be in the expansion phase. 

For the assumption of homogeneity and isotropy, only the radial coordinate
can be taken into account. 

In spherical coordinates equations (\ref{eq: accelerazione mareale lungo x}),
(\ref{eq: accelerazione mareale lungo y}) and (\ref{eq: accelerazione mareale lungo z})
give for the radial coordinate, i.e. the distance $D$\begin{equation}
\ddot{D}=\frac{1}{2}\ddot{h}_{R}D.\label{eq: accelerazione mareale lungo r}\end{equation}

Equivalently we can say that there is a gravitational potential \begin{equation}
V(\overrightarrow{D},t)=-\frac{1}{4}\ddot{h}_{R}(t)D^{2},\label{eq:potenziale in gauge Lorentziana}\end{equation}

which generates the tidal forces, and that the motion of the test
masses is governed by the Newtonian equation

\begin{equation}
\ddot{\overrightarrow{r}}=-\bigtriangledown V.\label{eq: Newtoniana}\end{equation}

Because we are in the linearized theory, the solution of eq. (\ref{eq: accelerazione mareale lungo r})
can be found by using the perturbation method \cite{key-16,key-17},
obtaining

\begin{equation}
D=D_{0}+\frac{1}{2}D_{0}h_{R}(t)=(1+\frac{1}{2}h_{R})D_{0}=a(t)D_{0}\label{eq:  distance}\end{equation}

Deriving this equation with respect the time we also get \begin{equation}
\frac{dD}{dt}=D_{0}\frac{da(t)}{dt}.\label{eq:  derivative}\end{equation}

Thus the Hubble law is obtained: \begin{equation}
\frac{1}{D}\frac{dD}{dt}=H_{0},\label{eq:  Hubble 2}\end{equation}

where 

\begin{equation}
H_{0}=(\frac{1}{a}\frac{da}{dt})_{0}.\label{eq:  const Hubble 2}\end{equation}

\subsection{The cosmological redshift}

Let us now consider another point of view. The conformal line element
(\ref{eq: metrica FRW}) can be putted in spherical coordinates, obtaining
\begin{equation}
ds^{2}=[1+h_{R}(t)](dt^{2}-dr^{2}).\label{eq: metrica puramente scalare radiale}\end{equation}

In this line element the angular coordinates have been neglected because
of the assumption of homogeneity and isotropy. The condition of null
geodesic in the above equation gives \begin{equation}
dt^{2}=dr^{2}.\label{eq: metrica puramente piu' di Corda lungo x 2}\end{equation}

Thus, from eq. (\ref{eq: metrica puramente piu' di Corda lungo x 2}),
it results that the coordinate velocity of the photon in the gauge
(\ref{eq: metrica puramente scalare radiale}) is equal to the speed
of light. This because in the coordinates (\ref{eq: metrica puramente scalare radiale})
$t$ is only a time coordinate. The rate $d\tau$ of the proper time
(distance) is related to the rate $dt$ of the time coordinate from
\cite{key-16}

\begin{equation}
d\tau^{2}=g_{00}dt^{2}.\label{eq: relazione temporale}\end{equation}
From eq. (\ref{eq: metrica puramente scalare radiale}) it is $g_{00}=(1+h_{R})$.
Then, using eq. (\ref{eq: metrica puramente piu' di Corda lungo x 2}),
we obtain

\begin{equation}
d\tau^{2}=(1+h_{R}))dr^{2},\label{eq: relazione spazial-temporale}\end{equation}

which gives 

\begin{equation}
d\tau=\pm[(1+h_{R})]^{\frac{1}{2}}dr\simeq\pm[(1+\frac{1}{2}h_{R})]dr.\label{eq: relazione temporale 2}\end{equation}

We assume that that photons are travelling by the galaxy to Earth
in this case too, thus the negative sign is needed. 

Integrating this equation it is\begin{equation}
\int_{\tau_{1}}^{\tau_{0}}\frac{d\tau}{1+\frac{1}{2}h_{R}(t)}=\int_{r_{g}}^{0}dr=r_{g},\label{eq: redshift 1}\end{equation}

where $\tau_{1}$ and $\tau_{0}$ are the emission and reception istants
of the photon from galaxy and Earth respectively. If the light is
emitted with a delay $\bigtriangleup\tau_{1},$ it arrives on Earth
with a delay $\bigtriangleup\tau_{0}$. In this way \begin{equation}
\int_{\tau_{1}}^{\tau_{0}}\frac{d\tau}{1+\frac{1}{2}h_{R}(t)}=\int_{\tau_{1}+\bigtriangleup\tau_{1}}^{\tau_{0}+\bigtriangleup\tau_{0}}\frac{d\tau}{1+\frac{1}{2}h_{R}(t)}=r_{g}.\label{eq: redshift 2}\end{equation}

The radial coordinate $r_{g}$ is \textit{comoving} (i.e. constant
in the gauge (\ref{eq: metrica puramente scalare radiale})) because
the assumption of homogenity and isotropy implies $\partial_{z}h_{R}=0,$
which removes the $z$ dependence in the line element (\ref{eq: metrica FRW}).
Thus the only dependence in the line element (\ref{eq: metrica puramente scalare radiale})
is the $t$ dependence in the scale factor $a=1+\frac{1}{2}h_{R}(t)$.
Then, from equation (\ref{eq: redshift 2}) it is \begin{equation}
\begin{array}{c}
\int_{\tau_{1}}^{\tau_{0}}\frac{d\tau}{1+\frac{1}{2}h_{R}(t)}=\int_{\tau_{1}}^{\tau_{0}}\frac{d\tau}{1+\frac{1}{2}h_{R}(t)}+\\
\\+\int_{\tau_{1}}^{\tau_{0}+\bigtriangleup\tau_{0}}\frac{d\tau}{1+\frac{1}{2}h_{R}(t)}-\int_{\tau_{0}}^{\tau_{1}+\bigtriangleup\tau_{1}}\frac{d\tau}{1+\frac{1}{2}h_{R}(t)},\end{array}\label{eq: redshift 3}\end{equation}

which gives \begin{equation}
\int_{\tau_{1}}^{\tau_{0}+\bigtriangleup\tau_{0}}\frac{d\tau}{1+\frac{1}{2}h_{R}(t)}=\int_{\tau_{0}}^{\tau_{1}+\bigtriangleup\tau_{1}}\frac{d\tau}{1+\frac{1}{2}h_{R}(t)}.\label{eq: redshift 4}\end{equation}

This equation can be semplified, obtaining \begin{equation}
\int_{0}^{\bigtriangleup\tau_{0}}\frac{d\tau}{1+\frac{1}{2}h_{R}(t)}=\int_{0}^{\bigtriangleup\tau_{1}}\frac{d\tau}{1+\frac{1}{2}h_{R}(t)},\label{eq: redshift 4a}\end{equation}

which gives\begin{equation}
\frac{\bigtriangleup\tau_{0}}{1+\frac{1}{2}h_{R}(t_{0})}=\frac{\bigtriangleup\tau_{1}}{1+\frac{1}{2}h_{R}(t_{1})}.\label{eq: redshift 5}\end{equation}

Then \begin{equation}
\frac{\bigtriangleup\tau_{1}}{\bigtriangleup\tau_{0}}=\frac{1+\frac{1}{2}h_{R}(t_{1})}{1+\frac{1}{2}h_{R}(t_{0})}=\frac{a(t_{1})}{a(t_{0})}.\label{eq: redshift 6}\end{equation}

But frequencies are inversely proportional to times, thus \begin{equation}
\frac{f_{0}}{f_{1}}=\frac{\bigtriangleup\tau_{1}}{\bigtriangleup\tau_{0}}=\frac{1+\frac{1}{2}h_{R}(t_{1})}{1+\frac{1}{2}h_{R}(t_{0})}=\frac{a(t_{1})}{a(t_{0})}.\label{eq: redshift 7}\end{equation}

If one recalls the definition of the \textit{redshift parameter} \cite{key-16,key-17,key-18,key-19,key-20}
\textit{}\begin{equation}
z\equiv\frac{f_{1}-f_{0}}{f_{0}}=\frac{\bigtriangleup\tau_{0}-\bigtriangleup\tau_{1}}{\bigtriangleup\tau_{1}},\label{eq: z}\end{equation}

using equation (\ref{eq: redshift 6}), equation (\ref{eq: z}) gives

\begin{equation}
z=\frac{a(t_{0})}{a(t_{1})}-1,\label{eq: z 2}\end{equation}

which is well known in the literature \cite{key-16,key-17,key-18,key-19,key-20}.

Thus, we have shown that our model is fine-tuned with the Hubble law
and the cosmological redshift.

\section{Conclusions }

An oscillating Universe which arises by the linearized \textbf{$R^{2}$}
theory of gravity, which is the simplest among the $f(R)$ theories
of gravity, has been discussed, showing that some observative evidences,
like the cosmological redshift and the Hubble law, are fine-tuned
with the model. In this context Dark Energy is seen like a pure curvature
effect arising by the Ricci scalar.

\section{Acknowledgements }

I would like to thank Salvatore Capozziello, Mauro Francaviglia and
Maria Felicia De Laurentis for useful discussions on the topics of
this paper. I thank the referee for precious advices and comments
that allowed to improve this paper. The EGO consortium has also to
be thanked for the use of computing facilities

\end{document}